\documentclass{amsart}
\usepackage{colortbl}

\newtheorem{rem}{Remark}[section]

\begin{document}
%---------------------------------------------------------------
\title{Weakly nonlocal irreversible thermodynamics -
the Ginzburg-Landau equation}
\author{P. V\'an}
\address{
Budapest University of Technology and Economics\\
Department of Chemical Physics\\
1521 Budapest, Budafoki \'ut 8.}
\email{vpet@phyndi.fke.bme.hu}

% \date{\today}

%---------------------------------------------------------------

\begin{abstract}
The variational approach to weakly nonlocal thermodynamic
theories is critically revisited in the light of modern
nonequilibrium thermodynamics. The example of Ginzburg-Landau
equation is investigated in detail.
\end{abstract}
\dedicatory{Dedicated to prof. W. Muschik on the occasion of his
65th birthday.}

\maketitle

%----------------------------------------------------------------

\section{Introduction}

In the last decades there has been a continuous interest in
developing generalized classical continuum theories that are able
to describe nonlocal effects. One of the most important (and
popular) strategies is to develop so called {\em weakly nonlocal}
continuum theories that is to incorporate higher order space
derivatives into the governing equations of continuum physics.
The crucial point in this investigations is to clarify the
relation of the new equations to the Second Law of thermodynamics.

From that point of view the weakly nonlocal continuum theories
can be divided into two groups. The theories in the first group
take seriously the thermodynamic requirements and the estabilished
structure of continuum physics. We can call them {\em
thermodynamic weakly nonlocal continuum theories}. The theories
in the second group does not follow the structure of classical
continuum physics and we can call them {\em variational weakly
nonlocal continuum theories} according to the basic method of
equation construction. Examples of thermodynamic theories are the
so called gradient theory (of thermomechanics) developed by
Kosi\'nski and Valanis \cite{Kos96p,Val96a,Val98a}, the virtual
power considerations of Germain and Maugin \cite{Mau90a1}, the
multifield theory of Mariano \cite{MarAug98a}, a microstructure
theory based on the concept of microforce balance of Gurtin
\cite{Gur96a} and the investigations toward the weakly nonlocal
extension of extended thermodynamics by Lebon, Jou and coworkers
\cite{LebAta95a,LebAta98a,LebGre96a,LebAta97a}. Instead of
reviewing and criticizing the different approaches our general
remark is, that they usually introduce new and disputable new
concepts that seem to be too special to serve as foundation of a
general nonlocal thermodynamic theory.

The second group is started from the investigations and method
introduced by Ginzburg and Landau and constructs a set of
prototypical classical weakly nonlocal equations like the
Ginzburg-Landau equation and the Cahn-Hilliard equation (see e.g.
\cite{HohHal77a} and the references therein). A unified treatment
of weak nonlocality based on this variational approach appears in
the so called GENERIC scheme developed by Grmela and \"Ottinger
\cite{GrmOtt97a,OttGrm97a}.

The following table compares the different nonlocal theories with
theories modeling memory effects according to their basic
structural ingredient of nonlocality (in space or time): \vskip
0.1in \setlength{\extrarowheight}{2mm}
\begin{tabular}{|l|p{1.5in}|p{1.5in}|}\hline
    & {Space}& {Time} \\ \hline\hline
Strongly nonlocal & space integrals & memory functionals \\\hline
Weakly nonlocal & gradient dependent constitutive functions
    & rate dependent constitutive relations \\\hline
Relocalized &\centering{???} & {internal variables} \\\hline
\end{tabular}
\vskip 0.1in

The table is self explanatory, what we should observe is, that all
of the weakly nonlocal approaches can be found in the third row
of the second column, that is they introduce gradient dependent
constitutive quantities (thermodynamic potential, entropy
current, conductivity tensor, etc...) to generate the nonlocal
effects. It is conspicuous the lacking of the nonlocal
counterpart of internal variables.

The main purpose of this paper is to confront the variational
approach with the thermodynamic requirements on the example of
Ginzburg-Landau equation. In the following section we will review
the traditional variational derivation of the Ginzburg-Landau
equation. In the second section a Ginzburg-Landau like equation
is constructed from thermodynamic principles preserving the
compatibility with classical continuum theories. The derivation
generalizes the internal variables to nonlocal phenomena and based
on a general entropy current. With this method the  and the
nonlocal constitutive functions can be generated with the help of
a force-current system like in classical irreversible
thermodynamics \cite{Van01a2}. In this way we can eliminate
several ad-hoc assumptions of the previously mentioned
thermodynamic approaches. The nonlocality of internal variables is
generated by thermodynamic requirements and the structure of the
theory. In the fourth section a more rigorous derivation, based
on the Liu procedure, shows the applicability conditions of the
irreversible thermodynamic treatment. From that considerations we
can recognize that the previous heuristic thermodynamic
derivation is almost completely general under some
straightforward physical requirements.

\section{Variational derivation of the Ginzburg-Landau equation}

From a thermodynamic point of view the Ginzburg-Landau equation
seems to be the first nonlocal extension of an evolution equation
of an internal variable (order parameter, dynamic degree of
freedom, etc...). Its traditional derivation is based on the
introduction of a gradient dependent thermodynamic potential, a
so called (Helmholtz) free energy functional. In the most simple
case this thermodynamic potential can be written in the following
form
\begin{equation}
\mathcal{F}(\xi) = \int\left( f(\xi) + \frac{\lambda}{2}
 (\nabla\xi)^2 \right) {\rm d}V.
\label{HFree_fun}\end{equation}

\noindent where $\xi$ denotes the internal variable, and $f$ is
the 'equilibrium part' of the free energy. According to intuitive
thermodynamic requirements we may assume that the equilibrium of
the related physical system is characterized by the extremum
(minimum) of the above free energy functional. In this case we
perform a variation and get the 'functional derivative' of
$\mathcal{F}$ as a generalized intensive quantity conjugated to
the internal variable
$$
\delta_\xi\mathcal{F} = f'(\xi) - \lambda \Delta\xi.
$$

Here the dash denotes a derivative. We can get the {\em stationary
Ginzburg-Landau equation} assuming that this functional derivative
is zero
$$
f'(\xi) - \lambda \Delta\xi = 0.
$$

Introducing a 'relaxational dynamics' we can generate the time
dependent Ginz\-burg-Lan\-dau equation as
\begin{equation}
 \dot{\xi} = - l\delta_\xi\mathcal{F} =- l(f'(\xi) - \lambda
\Delta\xi).
\label{GinLan_eq}\end{equation}

\noindent where $l$ is a positive scalar coefficient (e.g. in case
of scalar internal variable and isotropic material). We may
introduce more general free energies, internal variables with
different tensorial order, nonisotropic material, etc ...; but the
essence of the variational derivations remains the same.

As it was pointed out by Gurtin, the most important problem with
that derivations is, that they have nothing to do with the
balances of the fundamental physical quantities. He emphasizes the
importance of the separation of balances from the constitutive
properties: "My view is that while derivations of the form ...
are useful and important, they should not be regarded as basic,
rather as precursors of more complete theories. While variational
derivations often point the way toward a correct statement of
basic laws, to me such derivations obscure the fundamental nature
of balance laws in any general framework that includes
dissipation." \cite{Gur96a}

There are three specific problems with the variational approach as
a result of the incompatibility with the structure of continuum
physics.

\begin{itemize}
\item The above derivation is nothing to do with a nonnegative
entropy production, e.g. the sign of $l$ is fixed according to
some stability requirements, not by pure thermodynamic conditions.
\item The whole equation is not derived from the variational
principle, the relaxational dynamics is an additional, independent
requirement.
\item The whole procedure restrict how the rate terms can
appear in the equation. Essentially they can be added by hand
waving physical arguments in each specific theory, but
experiments and theoretical considerations show well, that they
have a typical form, that is independent of the specific theory
and a generalized Ginzburg-Landau equation can be written as (see
\cite{Gur96a} and the references therein)
\begin{equation}
\dot{\xi} =- l(f'(\xi) - \lambda \Delta\xi) + k \Delta\dot{\xi}.
\label{GenGinLan_eq}\end{equation}

Here $k$ is a positive coefficient.
\end{itemize}

\section{Thermodynamic derivation of the Ginzburg-Landau equation}

In this case our task is to find an evolution equation of an
internal variable that corresponds to the requirement of
nonnegative entropy production. We suppose here that the entropy
function depends only on the internal variable and we will use the
notation $\Gamma_\xi := Ds({\xi})$. If $ {\xi}=0$ then
$\Gamma_\xi(0)= 0$, because $\xi$ is an internal,
dynamic variable. As regards the entropy current we apply the
previous physical assumptions also in this case: if $ {\xi}$ was
zero then there is no entropy flow. Moreover, in the light of
this assumption and according to the mean value theorem, the
entropy current can be written as a linear function of the
derivative of the entropy, as in classical irreversible
thermodynamics. The coefficient can depend on the internal
variable, too:
$$
{\bf j}_s(\xi) = {\bf A}(\xi) \Gamma_\xi.
$$

That form of the entropy current was suggested by
Ny\'\i{}ri \cite{Nyi91a1}. Therefore the entropy production
follows as
$$
\sigma_s = \dot{s}( {\xi}) + \nabla\cdot {\bf j}_s =
\Gamma_\xi(\dot{ {\xi}} + \nabla \cdot {\bf A}) + {\bf A}\cdot
\nabla \Gamma_\xi \geq 0.
$$

We can recognize a force-current structure. In isotropic materials
the two terms do not couple and the corresponding Onsagerian
equations in the linear approximation are
\begin{eqnarray}
\dot{ {\xi}} + \nabla\cdot {\bf A} &=& l_1 \Gamma_\xi,
\label{GLO-1}\\
{\bf A} &=& l_2 \nabla \Gamma_\xi.
\label{GLO-2}\end{eqnarray}

Eliminating {\bf A} from (\ref{GLO-1}) and (\ref{GLO-2}), we get
\begin{equation}
\dot{ {\xi}} = l_1 \Gamma_\xi - \nabla\cdot(l_2 \nabla
\Gamma_\xi). \label{Ginzburg-Landau}\end{equation}

Here we have got an equation that is similar to the
Ginzburg-Landau equation (\ref{GinLan_eq}). However, there are
some differences.
\begin{itemize}
\item At the second term of the right hand side, under the space
derivatives there is $\Gamma_\xi$  instead of $ {\xi}$. However,
$\Gamma_\xi$ is a homogeneous linear function of $ {\xi}$, being
an internal variable.
\item The sign of the material coefficients
is determined by the Second Law, not by loose stability
considerations.
\item There is no additional rate term at this level of
approximation.
\item To extend the derivation to nonlinear and anisotropic cases
is straightforward. However, the nonlinearities and anisotropies
show a different structure, than in the original equation.
\end{itemize}

This thermodynamic Ginz\-burg\--Lan\-dau equation was derived
also by Verh\'as under some slightly different assumptions, as a
governing equation for the transport of dynamic degrees of freedom
\cite{Ver83a,Ver97b}.

\subsection{Generalized thermodynamic Ginzburg-Landau equation}

We can get the generalized form of the Ginzburg-Landau equation
at the next level of approximation. A similar equation was
received by Gurtin with the principle of microforce balance. In
the thermodynamic derivation we consider the previously introduced
current intensity factor {\bf A} as an internal variable and
follow a similar procedure. For the sake of simplicity we
suppose, that the entropy function does not depend on {\bf A},
that is we are not interested in the associated memory effects,
we are investigating only the nonlocal extension. The form of the
entropy current is similar to that of the Cahn-Hilliard equation
\cite{Van01a2}:
$$
{\bf j}_s({\bf A}, \xi) = {\bf A} \Gamma_\xi + {\bf B}({\bf A},
\xi)\cdot {\bf A}.
$$

Here ${\bf B}$ is a second order tensor. This form is completely
general under the conditions of the mean value theorem, if we
exploit that there is no entropy flow when the internal variable
{\bf A} is zero. Now the entropy production will be
$$
\sigma_s = \Gamma_\xi(\dot{ {\xi}} + \nabla \cdot {\bf A}) + {\bf
A}\cdot (\nabla \Gamma_\xi + \nabla\cdot {\bf B}) + {\bf
B}:\nabla{\bf A} \geq 0.
$$

It is straightforward to put down the Onsagerian conductivity
equations, but after the elimination of {\bf B}, one cannot
simplify them further. Therefore, we  will treat here only the
simplest situation, when the material is isotropic and the
approximation is strictly linear (the conductivity coefficients
are constants). Now the conductivity equations are reduced to the
following form
\begin{eqnarray}
\dot{ {\xi}} + \nabla\cdot {\bf A} &=& l_1 \Gamma_\xi,
\label{gGLO-1}\\
{\bf A} &=& l_2 (\nabla \Gamma_\xi + \nabla \cdot {\bf B}),
\label{gGLO-2}\\
{\bf B} &=& l_3^1 \nabla\circ {\bf A} + l_3^2 (\nabla\circ {\bf
A})^* + l_3^3 \nabla\cdot {\bf A},
 \label{gGLO-3}\end{eqnarray}

\noindent where $l_1, l_2, l_3^1, l_3^2, l_3^3$ are positive,
scalar, constant coefficients. Now a simple calculation eliminates
{\bf A} and {\bf B} from the above equations and we get
\begin{equation}
\dot{ {\xi}} = l_1 \Gamma_\xi - l_2\Delta(1+l_1 l_3)\Gamma_\xi +
l_3 \Delta\dot{ {\xi}},
\end{equation}

\noindent where $l_3 = l_3^1 + l_3^2 + l_3^3$. The last term, that
is additional to (\ref{Ginzburg-Landau}) corresponds to the
generalized Ginzburg-Landau equation (\ref{GenGinLan_eq}). The
positivity (positive definiteness in a more general situation) of
the material coefficients is ensured by the Second Law. However,
we should observe, that this generalization has not changed the
the characteristic thermodynamic term which differs form the
original Ginzburg-Landau form: $\Gamma_\xi$ stands under the
Laplacian instead of $\xi$.

We can continue the introduction of new nonlocal internal
variables, putting  {\bf B} into the basic state space. In this
case ${\bf B}$ becomes internal variable and we can introduce a
corresponding current intensity factor. Continuing this
procedure, we can develop a whole phenomenological hierarchy of
weakly nonlocal transport equations of higher and higher orders.
The further research in this direction has a special importance
for the kinetic theories, because we do not have a well
established approximation scheme for nonlocal phenomena like the
momentum series expansion in case of memory effects. The outlined
phenomenological hierarchy of nonlocal equations can suggest a
similar approach for the kinetic equations including the
sensitive question regarding the closure of the corresponding
relations \cite{Lib90b,Net93a}. It is straightforward to extend
the above treatment considering memory and nonlocal effects
together.

\section{A more exact derivation of the Ginzburg-Landau equation}

There are some basic problems in the heuristic approach of
irreversible thermodynamics that should be addressed in a modern
treatment. We can avoid them applying a more rigorous form and
exploitation of the Second Law (nonnegative entropy production)
that was applied in the previous section. In this section we make
a distinction between the state variables and the constitutive
quantities at the beginning and we apply the Liu procedure
\cite{Liu72a,MusAta00a2}.

As we have seen in the heuristic treatment, the thermodynamic
Ginzburg-Landau equation was received as a general evolution
equation for an arbitrary internal variable with the requirement
of the compatibility with the Second Law. Therefore, let us denote
our basic state space spanned by an internal variable ${\xi}$
with $Z_\xi$.  We are to find an evolution equation of the
internal variable in the form:
\begin{equation}
\partial_t{ {\xi}} + \mathcal{F} = {\bf 0},
\label{EvXi}\end{equation}

\noindent with the requirement of a nonnegative entropy production
$$
\partial_t{s} + \nabla\cdot{\bf j}_s \geq 0.
$$

Here ${\bf j}_s$ is the entropy current and $\partial_t$ denotes
the partial time derivative. With the different notation of the
time derivatives we emphasize that in case of moving media some
further considerations are necessary. The constitution space,
where the constitutive quantities are defined is spanned by $
{\xi}$ and its first and second gradients $( {\xi}, \nabla
{\xi},$ $\nabla^2 {\xi})$. Therefore $C_{I} = Z_\xi \times
Lin(Z_\xi,\mathbb{R}^3)\times Bilin(Z_\xi,\mathbb{R}^3)$ is the
constitution space of the nonlocal dynamics of an arbitrary
scalar internal variable. The constitutive quantities are the
entropy, the entropy current and the form of the evolution
equation $(s,{\bf j}_s, \mathcal{F})$. Therefore in the Liu
procedure the nonnegative entropy production is supplemented by
(\ref{EvXi})
\begin{eqnarray*}
\partial_t{\xi} + \mathcal{F} &=& 0,\\
\partial_1s \partial_t  {\xi} &+&
    \partial_2s\partial_t\nabla {\xi} +
    \partial_3s\partial_t\nabla^2 {\xi} +
    \partial_1 {\bf j}_s \cdot \nabla {\xi} +
    \partial_2 {\bf j}_s : \nabla^2 {\xi} +
    \partial_3 {\bf j}_s \cdot :\nabla^3 {\xi} \geq 0.
\end{eqnarray*}

According to Liu's theorem there exist a $\Gamma$, to be
determined from the Liu equations, which can be written in a
particularly simple form
\begin{eqnarray*}
\partial_1 s - \Gamma &=& 0, \\
\partial_2 s &=& 0, \\
\partial_3 s &=& 0, \\
\partial_3 {\bf j}_s &=& 0.
\end{eqnarray*}

The dissipation inequality in our case is
\begin{equation}
\partial_1 {\bf j}_s \cdot\nabla  {\xi} +
\partial_2 {\bf j}_s :\nabla^2  {\xi}- \Gamma \mathcal{F} \geq 0.
\label{Dis-GLeq}\end{equation}

The solution of the Liu equations gives that $s$ depends only on
the internal variable $ {\xi}$, $\Gamma = \Gamma_\xi = Ds(\xi)$
is the derivative of the entropy and ${\bf j}_s$ does not depend
on the second gradient of $ {\xi}$. Unfortunately these
considerations do not simplify the entropy inequality at all, we
may look additional conditions. We write the entropy current  in
the following form
\begin{equation}
{\bf j}_s( {\xi},\nabla {\xi}) = {\bf A}(
{\xi},\nabla {\xi}) \Gamma( {\xi}).
\label{Nyiri_fu}
\end{equation}

Considering the Liu equations we can see again, that this form of
the entropy current is completely general. Under some
differentiability conditions the entropy inequality turns out to
have the same form that we received with the heuristic
considerations of the previous section
$$
{\bf A}\cdot \nabla\Gamma + (\nabla\cdot{\bf A} +
\mathcal{F})\Gamma \geq 0.
$$
Let us remark, that in (\ref{Dis-GLeq}) there are two constitutive
quantities (the entropy current and $\mathcal{F}$) and three
additive terms. To simplify the inequality for constructing a
force current system there is no other choice that we have done
here: we should unite two of the terms with some reasonable
physical assumption. Assumption (\ref{Nyiri_fu}) is almost purely
mathematical, the physical condition is hidden in the
differentiability of {\bf A}. But that is necessary for the
construction of constitutive functions, because, if {\bf A} is
continuously differentiable we can solve the resulted inequality
and get the Onsagerian structure.

\begin{rem}
In this procedure we applied the Coleman-Mizel form of the Second
Law, with the requirement that the nonnegative entropy production
is a consequence of pure material properties, and valid for all
(continuously) solutions of the evolution equation on the
internal variable \cite{MusEhr96a}.
\end{rem}

Let us observe that one of the consequences is that a gradient
dependent thermodynamic potential (entropy of free energy) is not
compatible with a nonnegative entropy production and a
relaxational (or any) kind of evolution equation for the internal
variable.

\section{Discussion and further remarks}

The applied thermodynamic procedure can be applied to receive
several other classical weakly nonlocal equations of continuum
physics, e.g. the Guyer-Krumhansl equation for nonlocal heat
conduction of the Cahn-Hilliard equation for first order phase
transitions \cite{Van01m}. The Liu procedure shows well the
conditions (that are mostly mathematical).

The differences between the traditional and the thermodynamic
form makes possible to compare the consequences of the equations
experimentally. Understanding the (space-time) structure
generating properties of the original Ginzburg-Landau equation
from a thermodynamic point of view seem to be really interesting.
Furthermore, we have seen that the experimentally observed
additional rate dependent term of the generalized Ginzburg-Landau
equation is a natural consequence of the thermodynamic approach.
More thorough considerations on the possibilities of variational
principles show well that it would be hard to force that rate
dependent terms into a variational principle (see e.g.
\cite{VanMus95a,VanNyi99a}).

\section{Acknowledgements}

This research was supported by OTKA T034715 and T034603.


\begin{thebibliography}{10}

\bibitem{Kos96p}
W.~Kosi\'nski.
\newblock A modified hyperbolic framwork for thermoelastic materials with
  damage.
\newblock In W.~Kosi\'nski, R~de~Boer, and D.~Gross, editors, {\em Problems of
  environmental and damage Mechanics}, pages 157--172, Warszawa, 1997.
  IPPT-PAN.
\newblock Proceedings of SolMech'96 Conference, Mierki, Sept. 9-14. 1996,
  Poland.

\bibitem{Val96a}
K.~C. Valanis.
\newblock A gradient theory of internal variables.
\newblock {\em Acta Mechanica}, 116:1--14, 1996.

\bibitem{Val98a}
K.~C. Valanis.
\newblock A gradient thermodynamic theory of self-organization.
\newblock {\em Acta Mechanica}, 127:1--23, 1998.

\bibitem{Mau90a1}
G.~A. Maugin.
\newblock Internal variables and dissipative structures.
\newblock {\em Journal of Non-Equilibrium Thermodynamics}, 15:173--192, 1990.

\bibitem{MarAug98a}
P.~M. Mariano and G.~Augusti.
\newblock Multifield description of microcracked continua: A local model.
\newblock {\em Mathematics and Mechanics of Solids}, 3:183--200, 1998.

\bibitem{Gur96a}
M.~G. Gurtin.
\newblock Generalized {G}inzburg-{L}andau and {C}ahn-{H}illiard equations based
  on a microforce balance.
\newblock {\em Physica D}, 92:178--192, 1996.

\bibitem{LebAta95a}
G.~Lebon, M.~Torrisi, and A.~Valentini.
\newblock A non-local thermodynamic analysis of second sound propagation in
  crystalline dielectrics.
\newblock {\em Journal of Physics: Condensed Matter}, 7:1461--1474, 1995.

\bibitem{LebAta98a}
G.~Lebon, D.~Jou, J.~Casas-V\'azquez, and W.~Muschik.
\newblock Weakly nonlocal and nonlinear heat transport in rigid solids.
\newblock {\em Journal of Non-Equilibrium Thermodynamics}, 23:176--191, 1998.

\bibitem{LebGre96a}
G.~Lebon and M.~Grmela.
\newblock Weakly nonlocal heat conduction in rigid solids.
\newblock {\em Physics Letters A}, 214:184--188, 1996.

\bibitem{LebAta97a}
G.~Lebon, D.~Jou, J.~Casas-V\'azquez, and W.~Muschik.
\newblock Heat conduction at low temperature: A non-linear generalization of
  the {G}uyer-{K}rumhansl equation.
\newblock {\em Periodica Polytechnica Chemical Engineering}, 41(2):185--196,
  1997.
\newblock Lecture held on 'Minisymposium on Non-Linear Thermodynamics and
  Reciprocal Relations', September 22-26, Balatonvil\'agos, Hungary.

\bibitem{HohHal77a}
P.~C. Hohenberg and B.~I. Halperin.
\newblock Theory of dynamic critical phenomena.
\newblock {\em Reviews of Modern Physics}, 49(3):435--479, 1977.

\bibitem{GrmOtt97a}
M.~Grmela and H.~C. \"Ottinger.
\newblock Dynamics and thermodynamics of complex fluids. {I}. development of a
  general formalism.
\newblock {\em Physical Review E}, 56(6):6620--6632, 1997.

\bibitem{OttGrm97a}
H.~C. \"Ottinger and M.~Grmela.
\newblock Dynamics and thermodynamics of complex fluids. {II}. illustrations of
  a general formalism.
\newblock {\em Physical Review E}, 56(6):6633--6655, 1997.

\bibitem{Van01a2}
P.~V\'an.
\newblock Weakly nonlocal irreversible thermodynamics - the {G}uyer-{K}rumhansl
  and the {C}ahn-{H}illiard equations.
\newblock {\em Physics Letters A}, 290(1-2):88--92, 2001.

\bibitem{Nyi91a1}
B.~Ny\'\i{}ri.
\newblock On the entropy current.
\newblock {\em Journal of Non-Equilibrium Thermodynamics}, 16:179--186, 1991.

\bibitem{Ver83a}
J.~Verh\'as.
\newblock On the entropy current.
\newblock {\em Journal of Non-Equilibrium Thermodynamics}, 8:201--206, 1983.

\bibitem{Ver97b}
J.~Verh\'as.
\newblock {\em Thermodynamics and {R}heology}.
\newblock Akad\'emiai Kiad\'o and Kluwer Academic Publisher, Budapest, 1997.

\bibitem{Lib90b}
R.~L. Liboff.
\newblock {\em Kinetic Theory (Classical, Quantum, and Relativistic
  Descriptions}.
\newblock Prentice Hall, Englewood Cliffs, New Jersey, 1990.

\bibitem{Net93a}
R.~E. Nettleton.
\newblock Reciprocity and consistency in non-local {E}xtended {T}hermodynamics.
\newblock {\em Open Systems and Information Dynamics}, 2(1):41--47, 1993.

\bibitem{Liu72a}
I-Shih Liu.
\newblock Method of {L}agrange multipliers for exploitation of the entropy
  principle.
\newblock {\em Archive of Rational Mechanics and Analysis}, 46:131--148, 1972.

\bibitem{MusAta00a2}
W.~Muschik, C.~Papenfuss, and H.~Ehrentraut.
\newblock A sketch of continuum thermodynamics.
\newblock {\em Journal of Non-Newtonian Fluid Mechanics}, 96(1-2):255--290,
  2001.

\bibitem{MusEhr96a}
W.~Muschik and H.~Ehrentraut.
\newblock An amendment to the {S}econd {L}aw.
\newblock {\em Journal of Non-Equilibrium Thermodynamics}, 21:175--192, 1996.

\bibitem{Van01m}
P.~V\'an.
\newblock Weakly nonlocal irreversible thermodynamics.
\newblock to be submitted to Physica D, 2001.

\bibitem{VanMus95a}
P.~V\'an and W~Muschik.
\newblock Structure of variational principles in nonequilibrium thermodynamics.
\newblock {\em Physical Review E}, 5(4):3584--3590, 1995.

\bibitem{VanNyi99a}
P.~V\'an and B.~Ny\'\i{}ri.
\newblock Hamilton formalism and variational principle construction.
\newblock {\em Annalen der Physik (Leipzig)}, 8:331--354, 1999.

\end{thebibliography}
\end{document}